\documentclass[10pt,conference,a4paper]{IEEEtran}

\usepackage{amsmath}
\usepackage{booktabs}
\usepackage{cite}
\usepackage[keeplastbox]{flushend}
\usepackage{graphicx}
\usepackage{icomma}
\usepackage[utf8]{inputenc}
\usepackage{multirow}
\usepackage[super]{nth}
\usepackage{setspace}
\usepackage[caption=false,font=footnotesize]{subfig}
\usepackage{tikz}
\usepackage{url}
\usetikzlibrary{arrows.meta, decorations.pathreplacing, fit, positioning}

\title{Estimation of Rate Control Parameters for Video Coding Using CNN}
\author{
{Maria Santamaria{\small\(^{1,*}\)}%
\thanks{\textcopyright 2019 IEEE. Personal use of this material is permitted. Permission from IEEE  must  be  obtained  for  all  other  uses,  in  any  current  or  future  media, including reprinting/republishing this material for advertising or promotional purposes, creating new collective works, for resale or redistribution to servers or lists, or reuse of any copyrighted component of this work in other works.}%
, Ebroul Izquierdo{\small\(^1\)}, Saverio Blasi{\small\(^2\)}, Marta Mrak{\small\(^2\)}}
\vspace{1.6mm}\\
\fontsize{10}{10}\selectfont\itshape
\(^1\)\,Multimedia and Vision Group, Queen Mary University of London, London, United Kingdom \\
\(^2\)\,British Broadcasting Corporation, London, United Kingdom \\
\fontsize{9}{9}\selectfont\ttfamily\upshape
\(^*\)\,m.santamariagomez@qmul.ac.uk
}

\IEEEoverridecommandlockouts
\begin{document}
\maketitle

\begin{abstract}
Rate-control is essential to ensure efficient video delivery. Typical rate-control algorithms rely on bit allocation strategies, to appropriately distribute bits among frames. As reference frames are essential for exploiting temporal redundancies, intra frames are usually assigned a larger portion of the available bits. In this paper, an accurate method to estimate number of bits and quality of intra frames is proposed, which can be used for bit allocation in a rate-control scheme. The algorithm is based on deep learning, where networks are trained using the original frames as inputs, while distortions and sizes of compressed frames after encoding are used as ground truths. Two approaches are proposed where either local or global distortions are predicted.
\vspace{0.2mm}\\
\end{abstract}


\section{Introduction}\label{sec:intro}
Modern video coding standards, such as the H.265/High Efficiency Video Coding (HEVC), make use of complex mechanisms to provide remarkable compression efficiency. For distribution, frames are encoded using so called random-access configurations, in which most frames are inter-predicted, while a few intra frames are inserted periodically in the sequence (the number of frames between two intra frames is referred to as the intra-period). Intra frame coding uses prediction to decrease spatial redundancies, transform coding of residual signals, quantisation, and entropy coding to reduce statistical redundancies~\cite{HEVC}. Due to the inherent complexity of these modules, it is generally difficult to estimate the effects of an encoder on a given frame in terms of the number of bits and the distortion without actually encoding it. Conversely, rate-control mechanisms typically work by allocating the available number of bits per second among the frames in an intra-period, and then appropriately setting parameters to meet this allocation. Allocating the correct number of bits for intra frames is crucial, since such frames typically need significantly more bits than inter frames (due to the reduced efficiency of the encoder scheme). However, they should also be encoded at the highest quality, as they are used for reference by subsequent inter frames~\cite{Wang2015}. As such, schemes to accurately predict the number of bits and distortion generated by an intra frame encoder are highly beneficial.

A method based on deep learning to estimate distortion and number of bits needed to encode an intra frame is proposed in this paper. A first CNN is modelled to estimate the compressed frame size, measured as bits-per-pixel (bpp), and the average distortion, measured using the Peak Signal-to-Noise Ratio (PSNR) between original and compressed frames, obtained using different Quantisation Parameters (QPs). An additional CNN is also proposed to estimate distortion maps, namely pixel-wise maps of absolute differences between original and reconstructed frames, which may be used for block-wise rate-control or adaptive-quantisation schemes. The CNN computes the maps based on the original frame and an input QP\@.

\section{Related Work}\label{sec:works}
Methods based on deep learning have been shown to be very successful in different estimation tasks. In particular, Convolutional Neural Networks (CNNs) have earned a lot of attention in recent years due to their good performance, and have been extensively used for classification and segmentation~\cite{Goodfellow2016}, super resolution~\cite{Chua2013}, noise removal~\cite{Eigen2013} or depth estimation~\cite{Eigen2014}. 

Deep learning has also been used in video coding for various applications, including: frame partitioning~\cite{Li2017}, intra mode selection~\cite{Laude2016}, arithmetic coding~\cite{Song2017}, compressed frame sizes-distortion modelling~\cite{Xu2017} and post processing~\cite{Zhou2018}. Laude and Ostermann~\cite{Laude2016} introduced a CNN-based classifier for intra mode decision. The CNN takes an input block, and outputs the predicted intra mode to be used. Training uses original samples to avoid dependencies on other encoder decisions and reconstructed data, allowing to process several blocks in parallel. Li \textit{et al.}~\cite{Li2017} proposed a learning-based classifier to determine the partitioning of coding tree units (CTUs). Three CNNs are modelled to learn the split decision of CTUs at different depth levels, following maximum and minimum CTU sizes on HEVC\@. Song \textit{et al.}~\cite{Song2017} introduced a two-fold CNN-based arithmetic coding. First, a CNN is used to predict the distribution of the intra modes taking as input the Most Probable Modes (MPMs) of the current block and reconstructed neighbouring blocks. Subsequently, the predicted distributions are used in a multi-level arithmetic coding engine. Zhou \textit{et al.}~\cite{Zhou2018} proposed a CNN to replace deblocking filter and Sample Adaptive Offset (SAO).

An approach was presented by Xu \textit{et al.}~\cite{Xu2017}, where CNNs are used to estimate distortion maps and compressed frame sizes. Firstly, distortion maps are calculated with respect to the Structural Similarity Index (SSIM) between the original frame and its reconstruction. Secondly, compressed frame sizes are estimated, in the form of a vector of bits obtained after encoding a frame using different QPs. Both CNNs only use linear activations and can therefore be modelled as a combination of linear functions. 

\section{Proposed Approach}\label{sec:proposed-approach}
The CNNs proposed in~\cite{Xu2017}, from here on referred to as ``base CNNs'', were used as the starting point for the work proposed in this paper. As opposite to SSIM as used in~\cite{Xu2017}, most video encoders rely on Mean Square Error (MSE) based distortions to perform encoder-side mode decisions. Additionally, due to the non-linearity of several of the encoder blocks, using only linear activations may not be sufficient to provide accurate estimates. Finally, when dealing with practical applications, there may be a need for obtaining a low-complex estimate of distortion and number of bits. As such, the approach proposed here is different from the base CNN in that it is capable of predicting MSE distortions (instead of SSIM values) and makes use of non-linear activation functions. Moreover, in addition to a methodology to obtain local distortion maps, an additional CNN is proposed here which can provide a low-complexity estimate of average distortions for the whole frame (referred to as global distortions) and number of bits for a variety of QPs, in a single pass. The estimate of such global distortions was found to be in fact more accurate than that of local distortions, as shown in the rest of this paper.

\subsection{Local estimation of distortion maps}
The estimation of distortion maps was performed using a CNN with two inputs. The first input is the original frame data \(\mathbf{I}\). Only the luminance is considered, namely a matrix of dimension \(W \times H\), which is then normalised as follows:
\begin{equation}
  \hat{I}(x, y) = \frac{ I(x, y) } { 2 ^ {n - 1} },
\end{equation}
where \( x \in \{0, 1, 2, \ldots, W - 1\} \) and \(y \in \{0, 1, 2, \ldots, H - 1\} \), \(n\) is the bitdepth of the source samples. In addition, a second input is also considered, which consists of a normalised map of QP values (with respect to the maximum QP value $\text{QP}_{max}$, which in HEVC is set to 51), \(\hat{\mathbf{Q}}\), of dimension \(W \times H\), obtained as:
\begin{equation}
  \hat{Q}(x, y) = \frac{\text{QP}}{\text{QP}_{max}}.
\end{equation}

\begin{figure}[t]
  \centering
  \resizebox{.36\textwidth}{!}{%
  \begin{tikzpicture}[node distance=.2cm, align=center, base/.style = {rectangle, draw=black, minimum width=2.2cm, minimum height=.15cm, text centered, font=\sffamily\scriptsize},
  inout/.style = {base, rounded corners=3pt},
  layer/.style = {base}]
    \node (concat1) [layer] {Concat1};
    \node (img) [inout, above right=of concat1]{\(\hat{\mathbf{I}}\)};
    \node (qp) [inout, above left=of concat1]{\(\hat{\mathbf{Q}}\)};
    \node (conv1) [layer, below=of concat1] {Conv1 (3, 3, 64)};
    \node (conv2) [layer, below=of conv1] {Conv2 (3, 3, 64)};
    \node (pool1) [layer, left=of conv2] {MaxPool1 (2, 2)};
    \node (conv3) [layer, below=of pool1] {Conv3 (3, 3, 64)};
    \node (conv4) [layer, below=of conv3] {Conv4 (3, 3, 64)};
    \node (pool2) [layer, left= of conv4] {MaxPool2 (2, 2)};
    \node (conv5) [layer, below=of pool2] {Conv5 (3, 3, 64)};
    \node (up1) [layer, below=of conv5] {UpSample1};
    \node (conv6) [layer, below=of up1] {Conv6 (3, 3, 64)};
    \node (conv7) [layer, below=of conv6] {Conv7 (3, 3, 64)};
    \node (concat2) [layer, right=of conv7] {Concat1};
    \node (up2) [layer, below=of concat2] {UpSample2};
    \node (conv8) [layer, below=of up2] {Conv8 (3, 3, 64)};
    \node (conv9) [layer, below=of conv8] {Conv9 (3, 3, 64)};
    \node (concat3) [layer, right=of conv9] {Concat2};
    \node (conv10) [layer, below=of concat3] {Convolve (5, 5, 1)};
    \node (sum) [layer, below=of conv10] {Summation};
    \node (output) [inout, below=of sum] {\(\mathbf{M}\)};
    \node (convLayer) [layer, left=of conv8] {ConvJ (M, N, K)};
    \node (convolution) [layer, below=of convLayer] {Convolve (M, N, K)};
    \node (prelu) [layer, below=of convolution] {PReLU};
    \node (descr) [font=\sffamily\scriptsize, below=of prelu] {M: kernel width\\N: kernel height\\K: \# of kernels};
    \node [fit=(convLayer)(convolution)(prelu)(descr), draw, gray, dashed]{};
    \begin{scope}[all/.style={width=2mm, length=2mm}]
      \draw[->] (img.south) |- (concat1.east);
      \draw[->] (img) |- (sum);
      \draw[->] (qp.south) |- (concat1.west);
      \draw[->] (concat1) -- (conv1);
      \draw[->] (conv1) -- (conv2);
      \draw[->] (conv2) -- (pool1);
      \draw[->] (conv2) -- (concat3);
      \draw[->] (pool1) -- (conv3);
      \draw[->] (conv3) -- (conv4);
      \draw[->] (conv4) -- (pool2);
      \draw[->] (conv4) -- (pool2);
      \draw[->] (conv4) -- (concat2);
      \draw[->] (pool2) -- (conv5);
      \draw[->] (conv5) -- (up1);
      \draw[->] (up1) -- (conv6);
      \draw[->] (conv6) -- (conv7);
      \draw[->] (conv7) -- (concat2);
      \draw[->] (concat2) -- (up2);
      \draw[->] (up2) -- (conv8);
      \draw[->] (conv8) -- (conv9);
      \draw[->] (conv9) -- (concat3);
      \draw[->] (concat3) -- (conv10);
      \draw[->] (conv10) -- (sum);
      \draw[->] (sum) -- (output);
      \draw[->] (convolution) -- (prelu);
      \draw[decorate, decoration={brace, mirror, amplitude=5pt}] (convLayer.south west) -- (convLayer.south east);
    \end{scope}
  \end{tikzpicture}%
  }
  \caption{CNN \(G\). Dash-lined square indicates there are convolutional layers followed by a PReLU function.}\label{fig:map-based-cnn}
\end{figure}
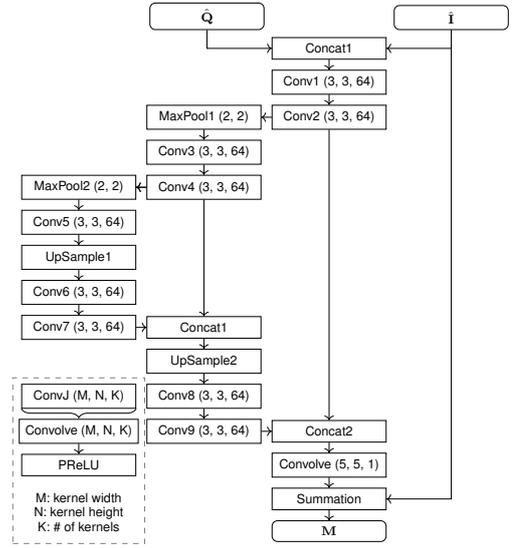

For the training, a set of ground truth distortion maps \(\mathbf{D}\) were used, namely sample-wise maps of absolute differences between the original and reconstructed frame. The goal of the network is to estimate the distortion map \(\mathbf{M} = G(\mathbf{\hat{I}}, \mathbf{\hat{Q}}) \approx \mathbf{D}\). As shown in Fig.~\ref{fig:map-based-cnn}, \(G\) is an CNN formed of residual connections, convolutions, non-linear mapping, down-sampling, up-sampling and skip connections.

\(G\) initially learns the differences between inputs and outputs, where such difference is modelled in the last layer as an element-wise summation between the output of the previous layer and \(\hat{\mathbf{I}}\). Secondly, convolutional layers use a stride of \(1 \times 1\), and filter sizes of \(3 \times 3\), except the final layer which uses a \(5 \times 5\) filter. Thirdly, non-linear mapping is achieved by adding Parametric Rectified Linear Unit (PReLU)~\cite{PReLU} after each convolutional layer, which increases the flexibility of the network. Max pooling layers adopt a filter size of \(2 \times 2\), the stride is \(1 \times 1\) and the output represents one quarter of the input. Up-sampling layers balance the size reduction introduced by max pooling layers. Finally, skip connections serve to aggregate multi-level features, which are modelled by concatenating the features learnt in the \nth{2} and \nth{4} convolutional layers with features learnt in \nth{9} and \nth{7} convolutional layers, respectively. 

The loss function used for training is the MSE:\@
\begin{equation}
  L_G = \frac{\sum_{x = 0}^{W - 1} \sum_{y = 0}^{H - 1} \big{(D(x, y) - M(x, y) \big)}^2}{W \cdot H}.
\end{equation}

\subsection{Estimation of number of bits and global distortions}
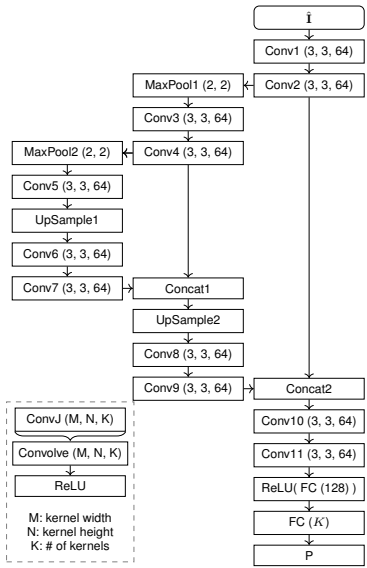
\begin{figure}[t]
  \centering
  \resizebox{.26\textwidth}{!}{%
  \begin{tikzpicture}[node distance=.2cm, align=center, base/.style = {rectangle, draw=black, minimum width=2.2cm, minimum height=.15cm, text centered, font=\sffamily\scriptsize},
  inout/.style = {base, rounded corners=3pt},
  layer/.style = {base}]
    \node (img) [inout]{\(\hat{\mathbf{I}}\)};
    \node (conv1) [layer, below=of img] {Conv1 (3, 3, 64)};
    \node (conv2) [layer, below=of conv1] {Conv2 (3, 3, 64)};
    \node (pool1) [layer, left=of conv2] {MaxPool1 (2, 2)};
    \node (conv3) [layer, below=of pool1] {Conv3 (3, 3, 64)};
    \node (conv4) [layer, below=of conv3] {Conv4 (3, 3, 64)};
    \node (pool2) [layer, left= of conv4] {MaxPool2 (2, 2)};
    \node (conv5) [layer, below=of pool2] {Conv5 (3, 3, 64)};
    \node (up1) [layer, below=of conv5] {UpSample1};
    \node (conv6) [layer, below=of up1] {Conv6 (3, 3, 64)};
    \node (conv7) [layer, below=of conv6] {Conv7 (3, 3, 64)};
    \node (concat1) [layer, right=of conv7] {Concat1};
    \node (up2) [layer, below=of concat1] {UpSample2};
    \node (conv8) [layer, below=of up2] {Conv8 (3, 3, 64)};
    \node (conv9) [layer, below=of conv8] {Conv9 (3, 3, 64)};
    \node (concat2) [layer, right=of conv9] {Concat2};
    \node (conv10) [layer, below=of concat2] {Conv10 (3, 3, 64)};
    \node (conv11) [layer, below=of conv10] {Conv11 (3, 3, 64)};
    \node (fc1) [layer, below=of conv11] {ReLU( FC (128) )};
    \node (fc2) [layer, below=of fc1] {FC (\(K\))};
    \node (output) [layer, below=of fc2] {P};
    \node (convLayer) [layer, below left=of conv9] {ConvJ (M, N, K)};
    \node (convolution) [layer, below=of convLayer] {Convolve (M, N, K)};
    \node (relu) [layer, below=of convolution] {ReLU};
    \node (descr) [font=\sffamily\scriptsize, below=of relu] {M: kernel width\\N: kernel height\\K: \# of kernels};
    \node [fit=(convLayer)(convolution)(relu)(descr), draw, gray, dashed]{};
    \begin{scope}[all/.style={width=2mm, length=2mm}]
      \draw[->] (img) -- (conv1);
      \draw[->] (conv1) -- (conv2);
      \draw[->] (conv2) -- (pool1);
      \draw[->] (conv2) -- (concat2);
      \draw[->] (pool1) -- (conv3);
      \draw[->] (conv3) -- (conv4);
      \draw[->] (conv4) -- (pool2);
      \draw[->] (conv4) -- (pool2);
      \draw[->] (conv4) -- (concat1);
      \draw[->] (pool2) -- (conv5);
      \draw[->] (conv5) -- (up1);
      \draw[->] (up1) -- (conv6);
      \draw[->] (conv6) -- (conv7);
      \draw[->] (conv7) -- (concat1);
      \draw[->] (concat1) -- (up2);
      \draw[->] (up2) -- (conv8);
      \draw[->] (conv8) -- (conv9);
      \draw[->] (conv9) -- (concat2);
      \draw[->] (concat2) -- (conv10);
      \draw[->] (conv10) -- (conv11);
      \draw[->] (conv11) -- (fc1);
      \draw[->] (fc1) -- (fc2);
      \draw[->] (fc2) -- (output);
      \draw[->] (convolution) -- (relu);
      \draw[decorate, decoration={brace, mirror, amplitude=5pt}] (convLayer.south west) -- (convLayer.south east);
    \end{scope}
  \end{tikzpicture}%
  }
  \caption{CNN \(F\). Dash-lined square indicates that each convolutional layer is followed by a ReLU function.}\label{fig:vector-based-cnn}
\end{figure}

An additional CNN was modelled to produce the estimate of the number of bits obtained with an HEVC encoder while intra coding a frame.  The CNN takes as input the normalised luminance image data \(\mathbf{\hat{I}}\), and is given ground truths in the form of a vector of scalars \(V\), where each element is the number of bits necessary to encode the frame with a certain QP value. A total \(K\) QP values are considered, and therefore \(K\) is the length of the vector. The goal is to estimate the vector \(P = F(\mathbf{\hat{I}}) \approx V \). As shown in Fig.~\ref{fig:vector-based-cnn}, the mapping \(F\) is a CNN similar to \(G\). Nevertheless, \(F\) uses Fully Connected (FC) layers that extract meaningful data from features. Moreover, convolutional layers are activated using Rectified Linear Unit (ReLU)~\cite{ReLU}, and the loss function \(L_F\) is the Mean Absolute Error (MAE):
\begin{equation}
  L_F = \frac{\sum_{i=1}^{K} |V(i) - P(i)|}{K}.
\end{equation}

In addition to being used for predicting the number of bits, the same CNN \(F\) was also trained  to predict global average distortions. In this case, each element in the the ground truths \(V\) is mean of the distortion map between original and reconstructed frame, as obtained when encoding with a given QP value.

The CNNs were trained using the parameters displayed in Table~\ref{tab:params}. The stop condition was defined in terms of epochs, where an epoch is defined as a complete training obtained by feeding all available samples in the training set to the network. In particular, the training was stopped in case the validation loss did not result in any improvement after additional 10 epochs of training. Furthermore, the loss functions were regularised by adding the \(\ell^2\)-norm of the training variables since on previous training/testing exercises better results were obtained with it.
\begin{table}[t]
  \centering
  \begin{scriptsize}
    \caption{Training parameters.}\label{tab:params}
    \begin{tabular}{c c c c}
      \toprule
      Batch size & Optimiser & Learning rate & Weight decay \\
      \midrule
      32 & Adam~\cite{adam} & 0.0001 & 0.0001 \\
      \bottomrule
    \end{tabular}
  \end{scriptsize}
\end{table}
\begin{table}[t]
  \centering
  \caption{Local correlation coefficients of distortion map estimates.}\label{tab:results-maps}
  \begin{scriptsize}
    \begin{tabular}{c c c c c c c}
      \toprule
      \multirow{2}{*}{CNN} & \multirow{2}{*}{Region} & \multicolumn{4}{c}{PCC} \\
      \cmidrule(lr){3-6}
        & & QP 22 & QP 27 & QP 32 & QP 37 \\
      \midrule
      \multirow{4}{*}{Base} & \(64 \times 64 \) & \( 0.58 \pm 0.5 \) & \( 0.57 \pm 0.6 \) & \( 0.79 \pm 0.4 \) & \( 0.86 \pm 0.3 \) \\
      & \(32 \times 32 \) & \( 0.54 \pm 0.4 \) & \( 0.59 \pm 0.4 \) & \( 0.71 \pm 0.3 \) & \(0.79 \pm 0.3 \) \\
      & \(16 \times 16 \) & \( 0.48 \pm 0.3 \) & \( 0.56 \pm 0.3 \) & \( 0.66 \pm 0.3 \) & \( 0.76 \pm 0.3 \) \\
      & \(8 \times 8 \) & \( 0.41 \pm 0.3 \) & \( 0.51 \pm 0.3 \) & \( 0.62 \pm 0.3 \) & \(0.72 \pm 0.2 \) \\
      \midrule
      \multirow{4}{*}{\(G\)} & \(64 \times 64 \) & \( 0.51 \pm 0.6 \) & \( 0.78 \pm 0.4 \) & \( 0.89 \pm 0.3 \) & \( 0.92 \pm 0.2 \) \\
      & \(32 \times 32 \) & \( 0.53 \pm 0.4 \) & \( 0.79 \pm 0.3 \) & \( 0.90 \pm 0.2 \) & \( 0.92 \pm 0.2 \) \\
      & \(16 \times 16 \) & \( 0.54 \pm 0.4 \) & \( 0.78 \pm 0.3 \) & \( 0.88 \pm 0.2 \) & \( 0.90 \pm 0.2 \) \\
      & \(8 \times 8 \) & \( 0.54 \pm 0.3 \) & \( 0.76 \pm 0.3 \) & \( 0.85 \pm 0.2 \) & \( 0.87 \pm 0.1 \) \\
      \bottomrule
    \end{tabular}
  \end{scriptsize}
\end{table}
\begin{figure}[t]
  \centering
  \setlength{\tabcolsep}{2pt}
  \begin{tabular}{c c c c c}
    \includegraphics[width=.08\textwidth]{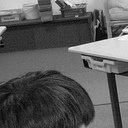} & \includegraphics[width=.08\textwidth]{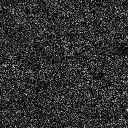} & \includegraphics[width=.08\textwidth]{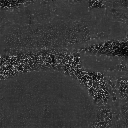} & \includegraphics[width=.08\textwidth]{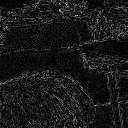} & \includegraphics[width=.08\textwidth]{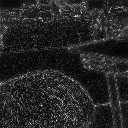} \\
    \includegraphics[width=.08\textwidth]{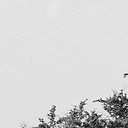} & \includegraphics[width=.08\textwidth]{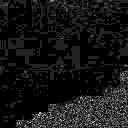} & \includegraphics[width=.08\textwidth]{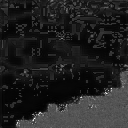} & \includegraphics[width=.08\textwidth]{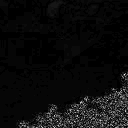} & \includegraphics[width=.08\textwidth]{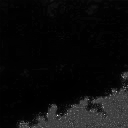} \\
    \footnotesize{Input} & \footnotesize{\(\mathbf{D}_{\text{QP}=22}\)} & \footnotesize{\(\mathbf{M}_{\text{QP}=22}\)} & \footnotesize{\(\mathbf{D}_{\text{QP}=37}\)} & \footnotesize{\(\mathbf{M}_{\text{QP}=37}\)} \\
  \end{tabular}
  \caption{Comparison of distortion maps. \(\mathbf{D}\) are the ground truths and \(\mathbf{M}\) are the estimates obtained using \(G\).}\label{fig:maps}
\end{figure}

\section{Experimental Results}\label{sec:evaluation}
The CNNs were implemented in TensorFlow and trained on an NVIDIA GeForce GTX 1080 GPU\@. MS COCO 2017~\cite{Lin2014} datasets are used for running the experiments: \(20,000\) frames are selected for training, \(5,000\) for validation and \(20,000\) for testing. The frames are cropped into \(128 \times 128\) patches and converted to YUV colour space. The HEVC reference software~\cite{HM} (HM 16.9) was used. Four different QPs were considered, namely \(22\), \(27\), \(32\) and \(37\).

The proposed methods are compared with the work in~\cite{Xu2017}. The base CNNs were implemented using the description provided within~\cite{Xu2017}, indicating the usage of linear activations for convolutional layers, training with Adam optimiser, learning rate of \(0.001\) and no regularisation. Furthermore, the training was done using a batch size of \(32\) and the same stop condition as in Section~\ref{sec:proposed-approach} was used. Additionally, the distortion is computed as the pixel-wise map of absolute differences, instead of SSIM, between original and reconstructed frames. While training the base CNNs, it was noticed that the networks would fluctuate around local minima without stabilising. This behaviour may be due to several factors, including the training dataset not being large enough or the variable updates using a too high learning rate. The proposed CNNs solve this issue by means of considering the regularisation within the loss function.

\begin{table}[t]
  \centering
  \begin{scriptsize}
    \caption{Comparison of compressed frame size estimates.}\label{tab:results-bpp}
    \begin{tabular}{c c c}
      \toprule
      Network & MAE & Fréchet distance \\
      \midrule
      Base & \( 10.454 \pm 9.346 \) & \( 15.558 \pm 16.748 \) \\
      \(F\) & \( 0.067 \pm 0.067 \) & \( 0.136 \pm 0.132 \) \\
      \bottomrule
    \end{tabular}
  \end{scriptsize}
\end{table}

\begin{table}[t]
  \centering
  \begin{scriptsize}
    \caption{Comparison of quality estimates.}\label{tab:results-psnr}
    \begin{tabular}{c c c}
      \toprule
      Network & MAE & Fréchet distance \\
      \midrule
      Base & \( 10.482 \pm 4.710 \) & \( 3.577 \pm 1.340 \) \\
      \(F\) & \( 0.757 \pm 0.719 \) & \( 1.260 \pm 0.896 \) \\
      \(G\) & \(1.216 \pm 0.359\) & \( 1.635 \pm 0.510 \) \\
      \bottomrule
    \end{tabular}
  \end{scriptsize}
  \vspace{-10pt}
\end{table}

Results obtained using the CNN \(G\) are presented here by measuring local correlation between the predicted \(\mathbf{M}\) and real \(\mathbf{D}\) distortion maps. Correlations were computed by squaring and averaging the distortion maps in blocks of different sizes. The values for each block were arranged in two vectors (one for the ground truth, and one for the estimated values, respectively), which were then compared using the Pearson Correlation Coefficient (PCC).

Table~\ref{tab:results-maps} shows a summary of the obtained PCC values in terms of QP and the size of the blocks. It can be noticed that the lower the QP, the lower the correlation between ground truths and estimates, indicating that the CNN predicts more easily in case of generically higher distortions (obtained with high QPs). Moreover, higher correlations are obtained when considering larger block sizes, which can be expected in that even in the case of local distortion estimates, the CNNs are more suitable for predicting global trends. This behaviour is confirmed through a visual comparison as exhibited in Fig.~\ref{fig:maps}. Although the estimated distortion map is not capable of estimating finer details in distortion present in the ground truth, trends in distortion variation are accurately estimated.
\begin{figure}[t]
  \centering
  {
  	\includegraphics[width=.2\textwidth]{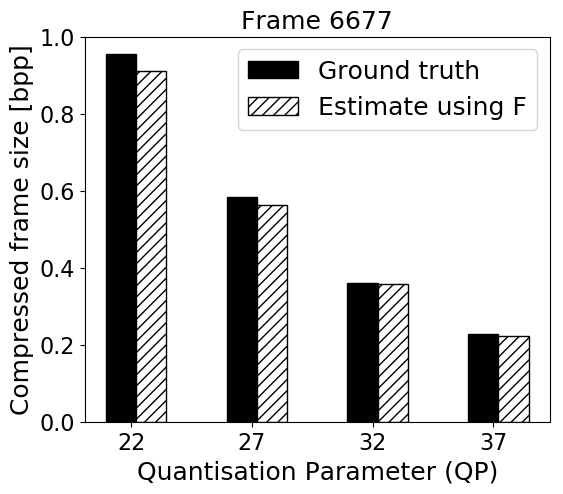}
  }
  {
  	\includegraphics[width=.2\textwidth]{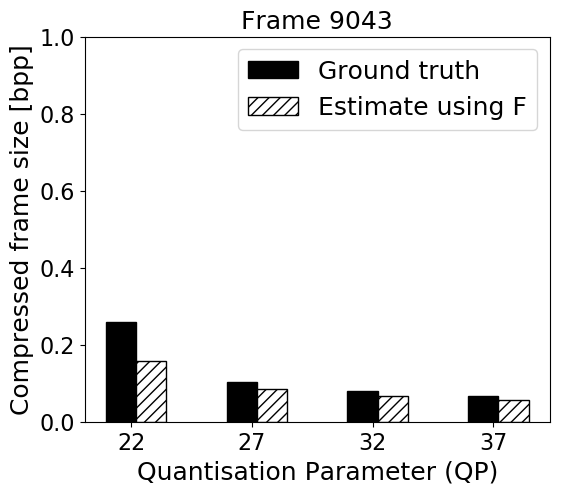}\label{fig:bpp-vs-qp-4}
  }
  \caption{Comparison of compressed frame sizes.}\label{fig:bpp-vs-qp}
  \vspace{-10pt}
\end{figure}
Results obtained using the CNN \(F\) are also presented both in terms of estimating global distortions and bits. These were analysed using the Fréchet distance~\cite{Wylie2013} (Euclidean), which measures similarity by calculating the minimum length of leash required to connect two curves. In this case, the distance between the interpolated curve of bpp or average PSNR values over QPs obtained using ground truth and estimations was computed. Tables~\ref{tab:results-bpp} and~\ref{tab:results-psnr} show these results, respectively. Average PSNR values are also reported for the \(G\) CNN.

\begin{figure}[t]
  \centering
  {
  	\includegraphics[width=.2\textwidth]{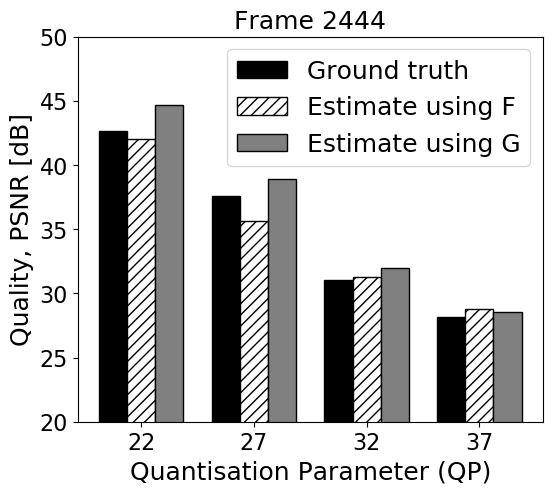}
  }
  {
  	\includegraphics[width=.2\textwidth]{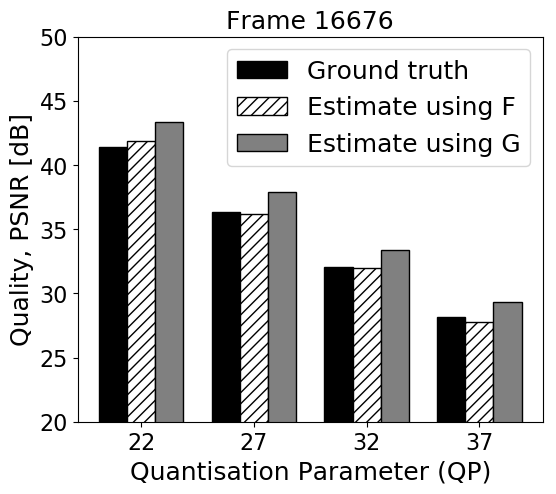}
  }
  \caption{Comparison of quality of reconstruction frames.}\label{fig:psnr-vs-qp}
  \vspace{-10pt}
\end{figure}

When considering estimate of bpp values, results show that the proposed network \(F\) outperforms the base model, since lower losses and lower Fréchet distances are obtained. Fig.~\ref{fig:bpp-vs-qp} displays bpp predictions per QP for two frames. Although difference can be seen in Fig.~\ref{fig:bpp-vs-qp-4}, there is a strong correlation between ground truths and predictions. Better results are obtained for higher QP values. Similarly, for distortion estimations, lower loss and lower Fréchet distance are obtained using the proposed networks. The predictions for two different frames are displayed in Fig.~\ref{fig:psnr-vs-qp}. In general, estimates obtained using \(F\) are better than those from \(G\), confirming that global estimations may be more suitable, unless the application requires local distortions to be available.

\section{Conclusions}\label{sec:conclusions}
This paper presents a CNN-based methodology to estimate distortion and number of bits obtained when intra coding original frames at different quality levels. One CNN is used to estimate vectors of compressed frame sizes or global distortions, whilst another CNN is used to estimate local distortion maps. Using the proposed methodology, these data can be estimated prior to the actual encoding process. Results show, in most cases, estimates are close and very correlated to real values. Future work includes the improvement of the CNNs, as well as the development of a complete bit allocation algorithm for rate-control applications.

\vspace{-2mm}
\section*{Acknowledgement}
\vspace{-0.5mm}
\begin{footnotesize}
The work leading to this paper was co-supported by the Engineering and Physical Sciences Research Council of the UK through an iCASE grant in cooperation with the British Broadcasting Corporation and by the project COGNITUS, which received funding from the European Union’s Horizon 2020 research and innovation programme under grant agreement No 687605. 
\end{footnotesize}

\bibliographystyle{IEEEtran}
\setstretch{0.9}
\bibliography{references}

\begin{thebibliography}{10}
\providecommand{\url}[1]{#1}
\csname url@samestyle\endcsname
\providecommand{\newblock}{\relax}
\providecommand{\bibinfo}[2]{#2}
\providecommand{\BIBentrySTDinterwordspacing}{\spaceskip=0pt\relax}
\providecommand{\BIBentryALTinterwordstretchfactor}{4}
\providecommand{\BIBentryALTinterwordspacing}{\spaceskip=\fontdimen2\font plus
\BIBentryALTinterwordstretchfactor\fontdimen3\font minus
  \fontdimen4\font\relax}
\providecommand{\BIBforeignlanguage}[2]{{%
\expandafter\ifx\csname l@#1\endcsname\relax
\typeout{** WARNING: IEEEtran.bst: No hyphenation pattern has been}%
\typeout{** loaded for the language `#1'. Using the pattern for}%
\typeout{** the default language instead.}%
\else
\language=\csname l@#1\endcsname
\fi
#2}}
\providecommand{\BIBdecl}{\relax}
\BIBdecl

\bibitem{HEVC}
G.~J. Sullivan, J.-R. Ohm, W.-J. Han, and T.~Wiegand, ``Overview of the high
  efficiency video coding ({HEVC}) standard,'' \emph{IEEE Transactions on
  Circuits and Systems for Video Technology}, vol.~22, no.~12, 2012.

\bibitem{Wang2015}
M.~Wang, K.~N. Ngan, and H.~Li, ``An efficient frame-content based intra frame
  rate control for high efficiency video coding,'' \emph{IEEE Signal Processing
  Letters}, vol.~22, no.~7, pp. 896--900, 2015.

\bibitem{Goodfellow2016}
I.~Goodfellow, Y.~Bengio, and A.~Courville, \emph{Deep Learning}.\hskip 1em
  plus 0.5em minus 0.4em\relax MIT Press, 2016.

\bibitem{Chua2013}
K.~K. Chua and Y.~H. Tay, ``Enhanced image super-resolution technique using
  convolutional neural network,'' in \emph{Advances in Visual Informatics},
  2013, pp. 157--164.

\bibitem{Eigen2013}
D.~Eigen, D.~Krishnan, and R.~Fergus, ``Restoring an image taken through a
  window covered with dirt or rain,'' in \emph{2013 IEEE International
  Conference on Computer Vision}, 2013, pp. 633--640.

\bibitem{Eigen2014}
D.~Eigen, C.~Puhrsch, and R.~Fergus, ``Depth map prediction from a single image
  using a multi-scale deep network,'' in \emph{Proceedings of the 27th
  International Conference on Neural Information Processing Systems - Volume
  2}, 2014, pp. 2366--2374.

\bibitem{Li2017}
T.~Li, M.~Xu, and X.~Deng, ``A deep convolutional neural network approach for
  complexity reduction on intra-mode {HEVC},'' in \emph{2017 IEEE International
  Conference on Multimedia and Expo (ICME)}, 2017.

\bibitem{Laude2016}
T.~Laude and J.~Ostermann, ``Deep learning-based intra prediction mode decision
  for {HEVC},'' in \emph{2016 Picture Coding Symposium (PCS)}, 2016.

\bibitem{Song2017}
R.~Song, D.~Liu, H.~Li, and F.~Wu, ``Neural network-based arithmetic coding of
  intra prediction modes in {HEVC},'' in \emph{2017 IEEE Visual Communications
  and Image Processing (VCIP)}, 2017.

\bibitem{Xu2017}
B.~Xu, X.~Pan, Y.~Zhou, Y.~Li, D.~Yang, and Z.~Chen, ``{CNN}-based
  rate-distortion modeling for {H.265/HEVC},'' in \emph{2017 IEEE Visual
  Communications and Image Processing (VCIP)}, 2017.

\bibitem{Zhou2018}
L.~Zhou, X.~Song, J.~Yao, L.~Wang, and F.~Chen, ``{JVET-IO022-v3}:
  Convolutional neural network filter ({CNNF}) for intra frame,'' Tech. Rep.,
  2018.

\bibitem{PReLU}
K.~He, X.~Zhang, S.~Ren, and J.~Sun, ``Delving deep into rectifiers: Surpassing
  human-level performance on imagenet classification,'' in \emph{2015 IEEE
  International Conference on Computer Vision (ICCV)}, 2015, pp. 1026--1034.

\bibitem{ReLU}
V.~Nair and G.~E. Hinton, ``Rectified linear units improve restricted boltzmann
  machines,'' in \emph{Proceedings of the 27th International Conference on
  Machine Learning (ICML-10)}, 2010, pp. 807--814.

\bibitem{adam}
D.~P. Kingma and J.~Ba, ``Adam: {A} method for stochastic optimization,'' in
  \emph{Proceedings of the 3rd International Conference on Learning
  Representations (ICLR)}, 2015.

\bibitem{Lin2014}
T.-Y. Lin, M.~Maire, S.~Belongie, J.~Hays, P.~Perona, D.~Ramanan,
  P.~Doll{\'a}r, and C.~L. Zitnick, ``Microsoft {COCO}: Common objects in
  context,'' in \emph{Computer Vision -- ECCV 2014}, 2014, pp. 740--755.

\bibitem{HM}
{Joint Collaborative Team on Video Coding (JCT-VC)}. {HEVC} test model
  reference software ({HM}). \url{https://hevc.hhi.fraunhofer.de}.

\bibitem{Wylie2013}
T.~R. Wylie, \emph{The Discrete Frechet Distance with Applications}.\hskip 1em
  plus 0.5em minus 0.4em\relax Montana State University, 2013.

\end{thebibliography}
\vspace{-0.5mm}

\flushend%

\end{document}